\documentclass{article}

\usepackage{arxiv}

\usepackage[utf8]{inputenc} 
\usepackage[T1]{fontenc}    
\usepackage{hyperref}       
\usepackage{url}            
\usepackage{booktabs}       
\usepackage{amsfonts}       
\usepackage{nicefrac}       
\usepackage{microtype}      
\usepackage{lipsum}
\usepackage{graphicx}
\usepackage{cite}
\usepackage{amsmath,amssymb,amsfonts}
\usepackage{algorithmic}
\usepackage{graphicx}
\usepackage{textcomp}
\usepackage{booktabs}
\usepackage{amsmath}
\usepackage{pdfpages}
\usepackage{hyperref}
\usepackage{amsthm}
\usepackage[ruled,vlined,linesnumbered,resetcount]{algorithm2e}
\theoremstyle{plain}

\theoremstyle{definition}

\graphicspath{ {./images/} }

\title{Sequence Preserving Network Traffic Generation}

\author{
 Sigal Shaked \\
 Dept. of Software and Information Systems Eng.\\ Ben-Gurion University of the Negev\\ Be'er Sheva, Israel\\
  \texttt{shaksi@post.bgu.ac.il} \\
   \And
 Amos Zamir \\
 Dept. of Software and Information Systems Eng.\\ Ben-Gurion University of the Negev\\ Be'er Sheva, Israel\\
  \And
 Roman Vainshtein \\
 Dept. of Software and Information Systems Eng.\\ Ben-Gurion University of the Negev\\ Be'er Sheva, Israel\\
  \And
   Moshe Unger \\
 Dept. of Software and Information Systems Eng.\\ Ben-Gurion University of the Negev\\ Be'er Sheva, Israel\\
  \And
    Lior Rokach \\
 Dept. of Software and Information Systems Eng.\\ Ben-Gurion University of the Negev\\ Be'er Sheva, Israel\\
  \And 
   Rami Puzis \\
 Dept. of Software and Information Systems Eng.\\ Ben-Gurion University of the Negev\\ Be'er Sheva, Israel\\
  \And 
  Bracha Shapira \\
 Dept. of Software and Information Systems Eng.\\ Ben-Gurion University of the Negev\\ Be'er Sheva, Israel\\
}

\begin{document}
\maketitle

\begin{abstract}
We present the Network Traffic Generator (NTG),
a framework for perturbing recorded network traffic with the purpose of generating diverse but realistic background traffic for network simulation and what-if analysis in enterprise environments.
The framework preserves many characteristics of the original traffic recorded in an enterprise, as well as sequences of network activities. Using the proposed framework, the original traffic flows are profiled using 200 cross-protocol features. The traffic is aggregated into flows of packets between IP pairs and clustered into groups of similar network activities. Sequences of network activities are then extracted. We examined two methods
for extracting sequences of activities: a Markov model and a neural language model. Finally, new traffic is generated using the extracted model. We developed a prototype of the framework
and conducted extensive experiments based on two real network traffic collections. Hypothesis testing was used to examine the difference between the distribution of original and generated features, showing that 30-100\% of the extracted features were preserved. Small differences between n-gram perplexities in
sequences of network activities in the original and generated traffic, indicate that sequences of network activities were well preserved.
\end{abstract}



\maketitle

\section{Introduction}
\label{sec:introduction}
The use of network traffic generators has been on the rise in recent years, mainly due to personal identifiable information (PII) privacy restrictions that limit the use of real data. Synthetic data can be thought of as "fake" data created from "real" data. The beauty of it comes from its foundation in real data and real distributions, which make it almost indistinguishable from the original data. Its momentum, in the context of privacy, stems from the fact that many times it is legally impossible to share real data, but in fact, anonymized data is insufficiently useful. At these moments, establishing a synthetic dataset may present the best solution of both worlds - data that can be shared and yet resembles the original data.

The "usefulness" of synthetic data has been validated by studies like  \cite{bellovin2019privacy,patki2016synthetic,choi2017generating,dutta2018simulated}. In this study we examine the use of synthetic data on another type of data, network traffic data. The main idea behind the proposed method is to train a generative model based on data from a limited number of consenting users, and then generate broader and more diverse traffic accordingly.

Generated traffic which preserves important features of the original traffic, including the distribution of different users, applications, and network properties, can be useful in many ways. Generated traffic can, for example, serve as an input for higher level analysis of traffic policing, flow classification, and anomaly detection. Network traffic generators are also critical in lab testing environments where they can be used for what-if analysis or to assess the behavior and performance of new systems and real network hardware in enterprise environments. A bandwidth measurement tool, for example, can be used with generated data that maintains the number of packets. Another use of generated traffic is in simulated environments where realistic background traffic is required. Without the ability to continuously create new test conditions (generated data) on the network, systems risk not being able to cope
 with unexpected behavior and perform poorly. Therefore, appropriate methods are needed to create scalable, adjustable, and representative network traffic.

Existing solutions \cite{TIRUMALA1999,Barford1998,Sommers2004,Vishwanath2009} have focused on generating statistically representative traffic, but to the best of our knowledge, there is no system that also preserves sequence patterns of network activities. To better understand the meaning of preserving network activity sequences, consider the following example: downloading a PDF from a certain site is usually performed after reading an email from a Gmail inbox. Even if the analysis of recorded traffic shows that this pattern is statistically significant, none of the existing methods for generating artificial traffic will preserve it. Therefore, an intrusion detection system (IDS) evaluated using the generated traffic may fail to identify a possible source of an attack and show weaker performance than it would on the original traffic.

This leads us to the main contribution of this study. We propose the Network Traffic Generator (NTG), a framework for perturbing recorded network traffic with the purpose of generating diverse but realistic background traffic for network simulation scenarios. We developed a prototype of the framework and examined its use for generating network traffic that preserves both the distribution of various traffic characteristics and sequence patterns of network activities. The NTG uses hundreds of features to maintain as many attributes of the original traffic as possible, enabling a robust solution that is not limited to a specific protocol or application. Packet flows are clustered to identify similar network activities, and machine learning techniques are then used to produce sequential patterns of network activities that will be retained in the traffic generated. To evaluate the new framework, extensive experiments were performed on real network traffic. The results show that the proposed framework successfully preserves network activity sequences and preserves other traffic characteristics impressively.
\section{Related Work}
\subsection{Traffic Generation}
Existing traffic generators focus primarily on maintaining statistical distributions of various network traffic characteristics, where three main approaches can be recognized.
Packet-level traffic generators like iPerf \cite{TIRUMALA1999} are based on packets' interarrival time and packet size. The size of each packet sent, as well as the elapsed time between subsequent packets, are selected by the user, usually by defining statistical distributions for each variable. These generators have been criticized for being inaccurate \cite{Botta2010}. Furthermore, they are used to test the performance and scalability of network tools, and the traffic they generate lacks the richness and diversity of packet streams observed within real organizational networks; these kinds of tools focus on performance tests rather than on reproducing the nature of the traffic.

Application-level traffic generators mimic the behavior of specific network applications in terms of the traffic they generate. For example, Surge \cite{Barford1998} is an Internet workload generator built by integrating distributions of features extracted from Internet usage. These tools typically focus on creating application-level request sequences that lead to network traffic with similar traffic statistics to those of real traffic within the modeled application. Although useful for assessing the behavior and performance of host systems, these tools recreate only one type of application traffic, not the variety of traffic viewed on the Web.

Flow-level traffic generators such as Harpoon \cite{Sommers2004}, recreate traffic flows (where flow is defined as a series of packets between a given IP / port pair) representative of those viewed on Internet traffic, based on eight distributional characteristics of TCP and UDP flows. Parameters for these distributions can be automatically extracted from data collected from live traffic. These features enable Harpoon to create statistical workloads that are independent of any specific application. Yet, statistics for just eight properties are preserved in the generated traffic. Frameworks like Swing \cite{Vishwanath2009} try to overcome this weakness by considering the interplay between multiple layers (users, sessions, connections, and network characteristics), so that more properties of the generated traffic can be better preserved. Yet, all of the abovementioned methods mainly preserve statistics of various traffic characteristics but lack the ability to preserve common sequences within the traffic.

Dutta et al. \cite{dutta2018simulated} present a framework for simulating user bots by imitating the actions of real users and demonstrate its successful implementation for intrusion detection. While their work is based on predefined rules, we will approach a similar task with machine learning techniques that will allow an automated extraction of real user behavior.
\subsection{Traffic Analysis}
The machine learning algorithms used for analyzing network traffic can be broadly divided into two categories, namely, classification (or supervised learning) \cite{Bekerman2015,Zhang2015} and clustering (or unsupervised learning) \cite{Erman2006,Wang2011}. Clusters have some key advantages such as eliminating requirements for fully labeled training datasets and the ability to discover hidden classes that may represent previously unknown applications. In this work, we apply clustering techniques to effectively identify sets of similar flows (or similar network activities).

There are three categories of traffic classification methods: port-based, payload-based, and flow statistics-based \cite{Nguyen2008,Zhang2015}. The traditional port-based method is based on standard port testing used by known applications. However, it is not always reliable, because not all current applications use standard ports. Some applications even obfuscate themselves by using well-defined ports of other applications. The payload-based method looks for the application signature in the payload of IP packets that can help prevent the problem of dynamic ports. Hence, it is most prevalent in industry products today. However, the payload-based method often fails with encrypted traffic. The limitations of port-based and payload-based analysis have motivated the use of the flow statistics-based method. Statistics-based attributes relate to traffic's statistical characteristics (e.g., flow duration, idle time, packets' interarrival time and length), without the need for deep packet testing. The argument is that traffic generated by different types of applications presents distinct characteristics. The most common way is to extract statistical features that represent traffic flows and then apply supervised or unsupervised learning techniques to classify them. 

We adopted the latter approach due to its widespread use in recent years. Our network feature engineering approach is largely based on \cite{Bekerman2015} which proposed a method for detecting malware using a set of 972 engineered network features that represent aspects of network traffic. These properties cross protocols and network layers, and refer to different observation resolutions (transaction, session, flow, and conversation).

\section{Methodology}
\begin{figure*}[htbp!]  
\centering    
  \includegraphics[width=0.9\linewidth]{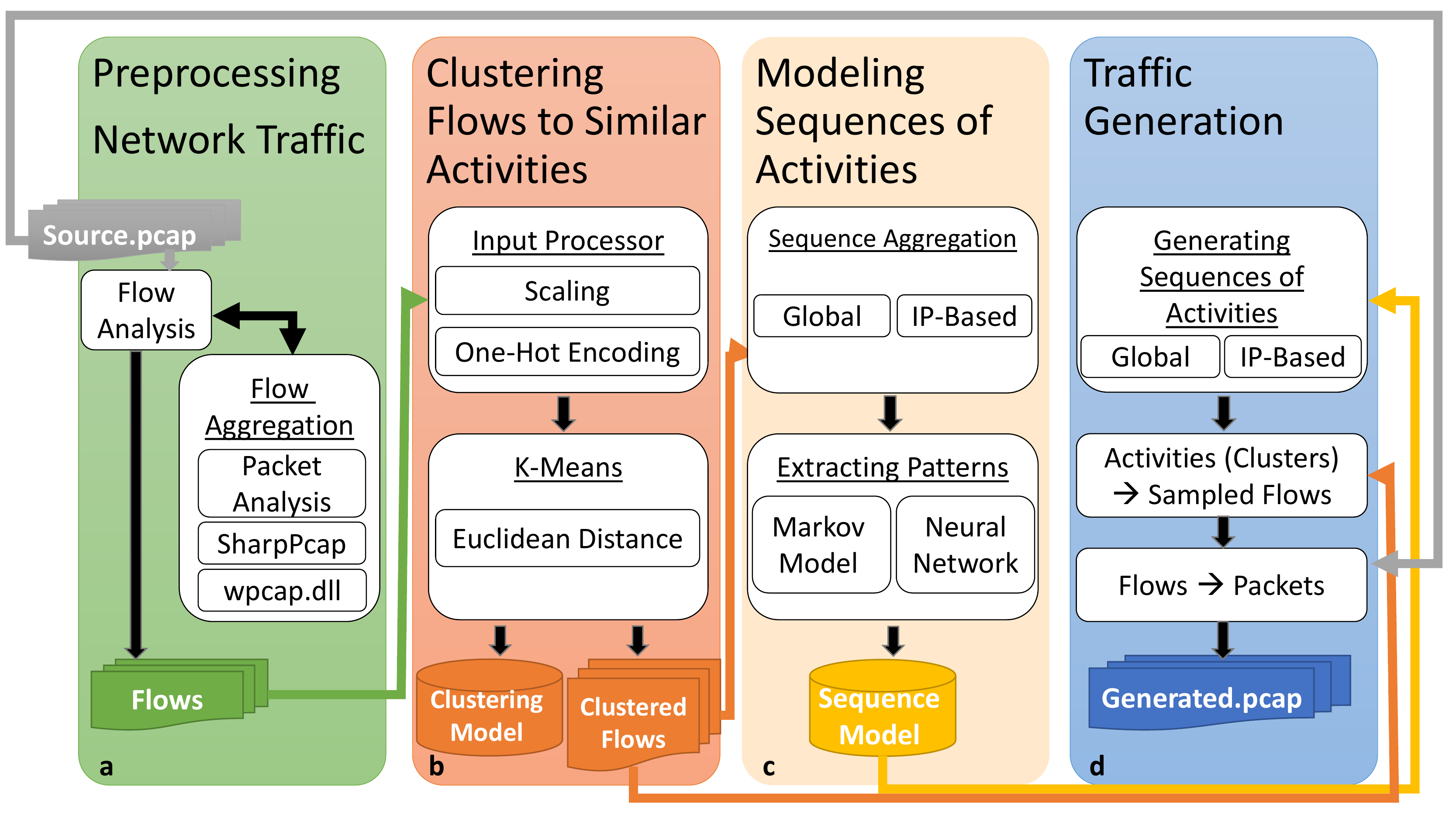}
  \caption[Data flow within the proposed framework]{Data flow within the proposed framework: (a) the original traffic is preprocessed into aggregated flows; (b) flows are clustered into groups of network activities; (c) a sequence model is trained to extract common patterns based on sequences of network activities (several methods are supported for this task); (d) new sequences of network activities are generated and then converted to sampled flows of packets that match each activity.}
  \label{fig:architecture}
\end{figure*}

The prototype of the NTG framework incorporates four phases comprised of interconnected sets of components as depicted in Fig~\ref{fig:architecture}: Preprocessing Network Traffic, Clustering Flows to Similar Activities, Modeling Sequences of Activities, and Traffic Generation.

In the preprocessing phase we focused on the two most common application layer protocols (DNS and HTTP). Nevertheless, the prototype can easily be extended to include additional protocols by extracting additional features. After preprocessing, the flow records are analyzed further in order to extract similar activities. In the next phase, sequences of activities are modeled using either the Markov model or neural networks, and finally, in the last phase the trained sequence model is utilized for generating artificial traffic.
\subsection{Preprocessing Network Traffic}
A basic entity considered in this study is the flow of traffic; this term refers to a sequence of packets that were passed between two peers during a single session, with the two partners being a unique 4-tuple consisting of source and destination IP addresses and ports. A TCP session starts with a successful handshake and ends with a timeout or packet with an RST or FIN flag from a peer. Since in this protocol two peers send packets in turn, two flows will be generated, one for each direction of the session (this distinction is necessary in order to learn a sequence of network activities per peer, instead of per two peers). The UDP session consists of all packets sent from client to server until a defined idle communication time or a maximal duration threshold is reached. A single flow is therefore generated for each session.

We identified 205 unique features that might shape the behavior of network traffic. The list of features is based on \cite{Bekerman2015}. We extracted features from three levels: 1) raw/packet level; 2) flow level and 3) application level.
The extracted features are detailed in Table~\ref{tab:features}. Raw/packet level features are extracted from each packet at the transportation layer level. We extracted fifteen TCP-based attributes and a single UDP-based feature. Each raw feature was transformed into a set of aggregative features that describe each feature within packets of a given flow. For example, “packet size” is extracted to eleven aggregative features like mean packet size (average number of packet size in the flow), packet size entropy, packet size first quartile, etc. 
Flow level features are naturally extracted from a flow. We extracted six flow level features; an example for such a feature is the number of packets within a flow. Application level features are features that appear in the application layer, they are only relevant to certain applications. HTTP user agent, for example, will be extracted for flows that suite HTTP applications. We extracted eight DNS-based features, eight HTTP-based features and seven SSL-based features.

Table~\ref{tab:features} shows several examples of features that appear in different protocols of network traffic. For each of the cumulative features we calculate the following statistics as additional features: minimum, first quartile, median, third quartile, maximum, mean, standard deviation, variance, and entropy.
\begin{table}[htbp!] 
\centering   
  \caption{Examples of features.}
  \label{tab:features}
  \begin{tabular}{cl}
    \toprule
    Level&Features\\
    \midrule
    Flow &time of day, packet interarrival time, the number of packets\\
    TCP &time to live, seq num, ack num\\
    UDP &checksum invalid\\
    DNS &additional records, canon names, response count\\
    HTTP &cookie, unique content types, bytes\\
    \bottomrule
  \end{tabular}
\end{table}
The preprocessor receives raw packets of captured network traffic as input and aggregates packets of the TCP and UDP transport layer protocols into flows. For each flow, a set of 205 features is extracted. In addition, some network level statistics are calculated, specifically the average time interval between flows and the number of flows for each pair of IPs.

In order to extract the features mentioned above, we developed a dedicated feature extraction component that processes the raw network traffic, divides it into flows, extracts the features, and provides the features as input to the clustering module. The preprocessing module is shown in Fig~\ref{fig:architecture}(a); it was implemented based on C\#'s SharpPcap library\footnote{\url{https://github.com/chmorgan/sharppcap}} to extract data from captured network traffic in the tcpdump format. The flow analysis component is aided by the packet parser to decode information at the packet level. It generates events and statistics and aggregates them into flow level features. In addition, for each flow, the flow analysis component aggregates references to each of the flow’s packets within the original tcpdump file and eventually outputs a set of network flows with the engineered features.
\subsection{Clustering Flows to Similar Activities}
 As a flow represents a particular network activity from start to finish, we can group similar activities into clusters that represent types of network activities. This step is needed in order to enable the production of sequence patterns of network activities. There are a variety of clustering algorithms available. The k-means algorithm \cite{Hartigan1979} was selected, because it is a quick and simple clustering algorithm. To measure the distance between flows, a distance measure must be selected. We adopt the traditional Euclidean distance, which is one of the most commonly used metrics for clustering problems  \cite{Erman2006}. Given a set of flows $F=\{f_1,..,f_n\},f_i \in R^M$, with $M$ features per flow, the distance between two flows is:
\begin{equation}
 dist(f_i,f_j )=\sqrt{\sum_{d=1}^{|M|}(f_i^d-f_j^d)^2}.
\end{equation}
The k-means algorithm divides the set of flows into $K$ disjoint clusters. For each cluster, the algorithm maximizes the homogeneity within the cluster by minimizing the sum of squared errors of the distance between each flow $f_j$ and the center of its cluster $c_i$: 
\begin{equation}
E= \sum_{i=1}^K \sum_{f_j \in c_i}^n(dist(c_i,f_j ))^2.
\end{equation}
The k-means algorithm starts with a random initialization of cluster centers and iterates towards the minimum error, where in each iteration flows are assigned to clusters with the nearest centers after which cluster centers are recalculated. Usually, the algorithm converges to stable clusters within a small number of iterations. We conducted some additional preprocessing steps in order to ensure that the set of flows is appropriate for the clustering task; this includes the conversion of nominal features to numerical features with the one-hot binarization method, as well as scaling the features to unit variance.

\subsection{Modeling Sequences of Network Activities}
Before designing our framework, we needed to consider various network activity sequence model aggregation options. In order to do this we were faced with some fundamental questions: What is the sequence on the Web that we want to maintain made of? Does it make sense to maintain a sequence of activities carried out by all network members, or should we keep sequences of activities made by pairs of communicating IPs (e.g., a sequence of activities from a source IP to a specific destination server)? We examine both of these network activity sequence model aggregation options. The first (global) refers to all of the traffic as a single sequence, and the second (IP-based) refers to each pair of communicating IP addresses as a separate sequence. In this research, we examine, among other things, which of the two aggregations is more suitable for our purposes.

  A sequence $s$ is an ordered list of network activities, represented by identifiers of clusters connected to flows that occurred in sequence: $s= \{ c_1 \rightarrow c_2 \rightarrow \ldots  \rightarrow c_{i} \}$, where $i \in K$. In the global aggregation, all of the traffic flows belong to the same sequence, whereas in the IP-based aggregation, flows belonging to the same sequence share source and destination IP addresses. The time of the first packet in each flow $t_i$ dictates the order of activities in the sequence. There is a transition between two sequential activities within the same sequence; the transition time between the i-th activity and the following activity is equal to $t_{i+1}-t_i$. Sequential consecutive states can be the same (for example:  $c_3 \rightarrow c_2  \rightarrow c_2$).

\begin{figure}[h!] 
\centering   
\includegraphics[width=\linewidth ]{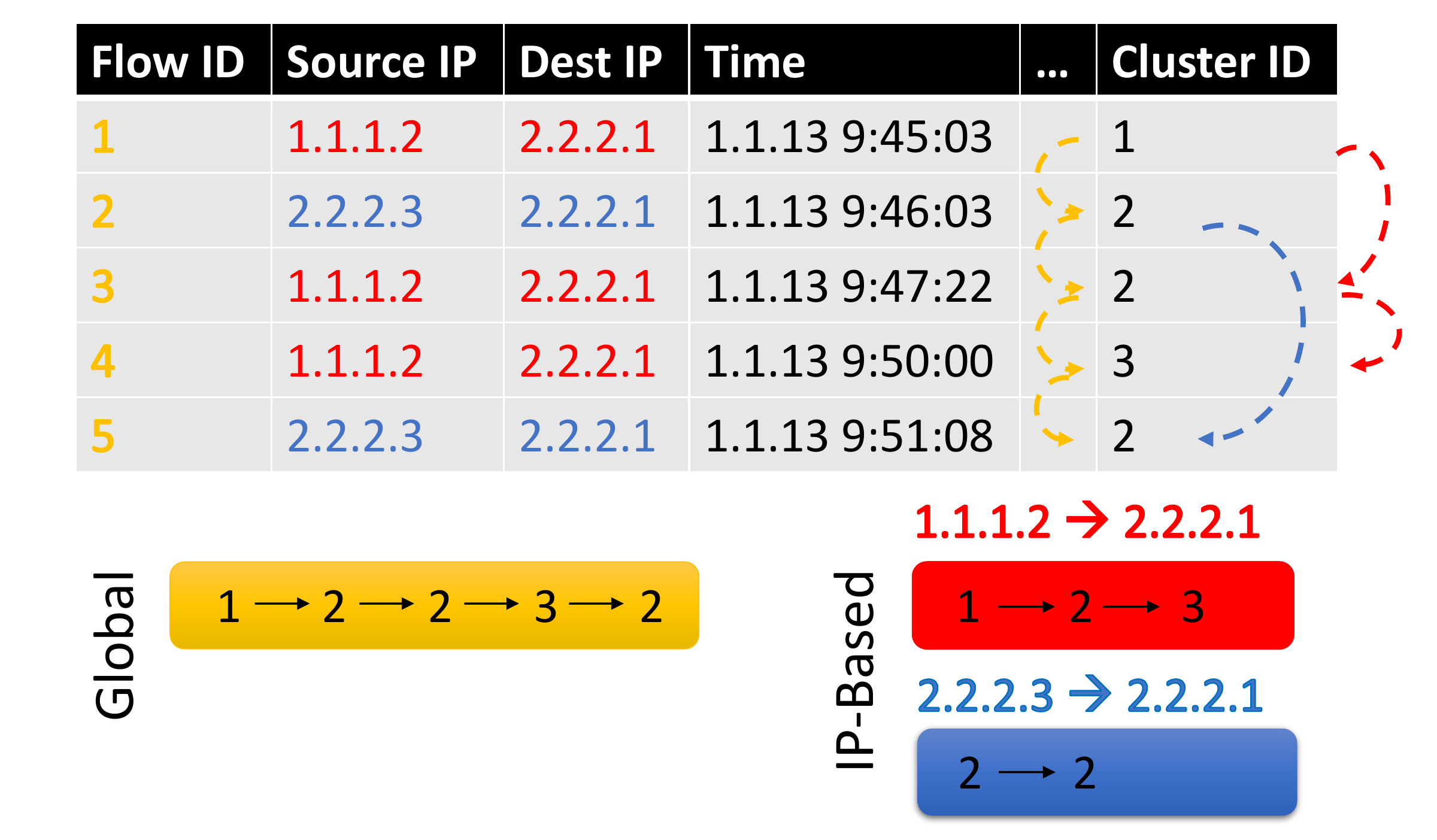}
\caption[Two sequence aggregations: global vs. IP-based.]{Two sequence aggregations: Global vs. IP-based. In the global aggregation (yellow) all flows in the traffic belong to the same sequence, while in the IP-based aggregation (red and blue), a sequence is created for each unique pair of source and destination IP addresses. The sequence is composed of cluster IDs attached to the relevant flows and is ordered by the time of the first packet in each flow.}
\label{fig:seqTopology}
\end{figure}
Fig~\ref{fig:seqTopology} illustrates the difference between the global and the IP-based sequence aggregations. In the global aggregation, all of the flows comprise one sequence (the yellow arrow indicates their order). In the IP-based aggregation, a sequence is generated for each unique pair of source and destination IP addresses. As there are two unique pairs in the example set, two sequences are produced: the first consists of three flows extracted from traffic that passed from 1.1.1.2 to 2.2.2.1 (marked in red), and the second consists of two flows that passed from 2.2.2.3 to 2.2.2.1 (marked in blue). In the IP-based aggregation, we add start and end states around each sequence to enable us to sample an initial state and predict the end of sequence (learning sequence sizes).

Searching for a strong algorithm for extracting a sequence model, in this study we examine two methods: the Markov model and the neural language model.
\subsubsection{Markov Models}
Markov models (MMs) \cite{Rabiner1986} are one of the most common probabilistic sequence models, and they have been applied to a wide variety of tasks, largely because of their ability to reduce complexity, particularly for long sequences. More specifically, an MM is a type of probabilistic graphical model (or Bayesian network) which has a chain topology. In a Bayesian network, each random variable is represented as a node in a graph, and the edges in the graph are directed and represent probabilistic dependencies between the random variables.

In an $m$-order Markov model, the probability for the appearance of a state in a sequence depends only on the previous $m$ states of the sequence. A second-order Markov model has the lowest computational cost, since it only examines the previous state when predicting the current state. The probability for the appearance of a state as the current state in a sequence according to a second-order Markov model is:
\begin{equation}
Pr(c_{i+1}=a | c_1,c_2, \ldots ,c_i ) \approx Pr(c_{i+1}=a | c_i ).
\end{equation}
In order to generate a sequence of clusters, frequencies of starting clusters and cluster transitions are collected; let $F_D$ denote the frequency of a given cluster according to a set of clustered flows $D$, so that: 
\begin{equation}
StartProb(a)=F_D(c_1=a),
\end{equation}
and:
\begin{equation}
TransitionProb(a,b)=\sum_{ \forall i} F_D(c_{i+1}=b | c_i=a).
\end{equation}
\subsubsection{Neural Language Model}
For the second examined method, we chose to rely on natural language processing (NLP), which deals with a different type of sequence. In the last decade, significant advances have been made in the area of textual sequences, while developing useful applications such as speech recognition, language translation, and information retrieval.

The objective of statistical language modeling is to learn the joint probability function of sequences of words in a language, or of sequences of network activities (that is, sequences of clusterIDs) in the context of the current study. The neural language model idea was proposed in \cite{Bengio2003}, where a neural network with an embedding layer was used for training word embeddings along with the parameters of the model. The neural language model simultaneously learns a distributed representation for each word (a word vector), along with the probability function of word sequences, which are expressed in terms of these representations.

When word vectors are used, each word is identified with a point in the vector space. The number of attributes (for example, $m = 64$ in our experiments) is much smaller than the size of the vocabulary. The probability function is expressed as a product of conditional probabilities of the next word given the previous words, (e.g., using a multilayer neural network to predict the next network activity given the previous activities, in our experiments). This function has parameters that can be iteratively tuned to maximize the log-likelihood of training data. The motivation for using word embeddings for discrete spaces is that when designing continuous variables, generalization can more easily be achieved, because the function to be learned can have some local smoothness properties.

A statistical language model can be represented by the conditional probability of the following word given each previous word:
\begin{equation}
\widehat{p}(w_1^T )= \prod_{t=1}^T \widehat{p}(w_t |w_1^{t-1}),\end{equation}
where $w_t$ is the t-th word, and writing subsequence $ w_i^j = (w_i\mathbin{,} w_{i+1}\mathbin{,}\cdots\mathbin{,} w_{j-1}\mathbin{,} w_j) $. The difficulty of this problem can be reduced by taking advantage of word order and the fact that words closer in the sequence are statistically more dependent. Therefore, we predict a word based on the context of the previous $n-1$ words:
\begin{equation}
\widehat{p}(w_t |w_1^{t-1} ) \approx \widehat{p}(w_t |w_{t-n+1}^{t-1} ).
\end{equation}
Typically, researchers have used trigrams ($n = 3$) and obtained state of the art results. In our experiments we used $n = 4$, to further enhance the model (as opposed to using bi-grams in the Markov model examined). Training such a large scale model is expensive. However, training large models which consider larger contexts, yield better results.

The training set is a sequence $w_1, \ldots, w_t$ of words $w_t \in V$, where the vocabulary $V$ is a large but finite set. The goal is to learn a good model $f(w_t,\ldots,w_{t-n+1} ) \approx p(w_t | w_{t-n+1}^{t-1} )$, in the sense that it gives high likelihood for the input sample. 
\begin{figure}[htbp!] 
\centering   
\includegraphics[width=\linewidth]{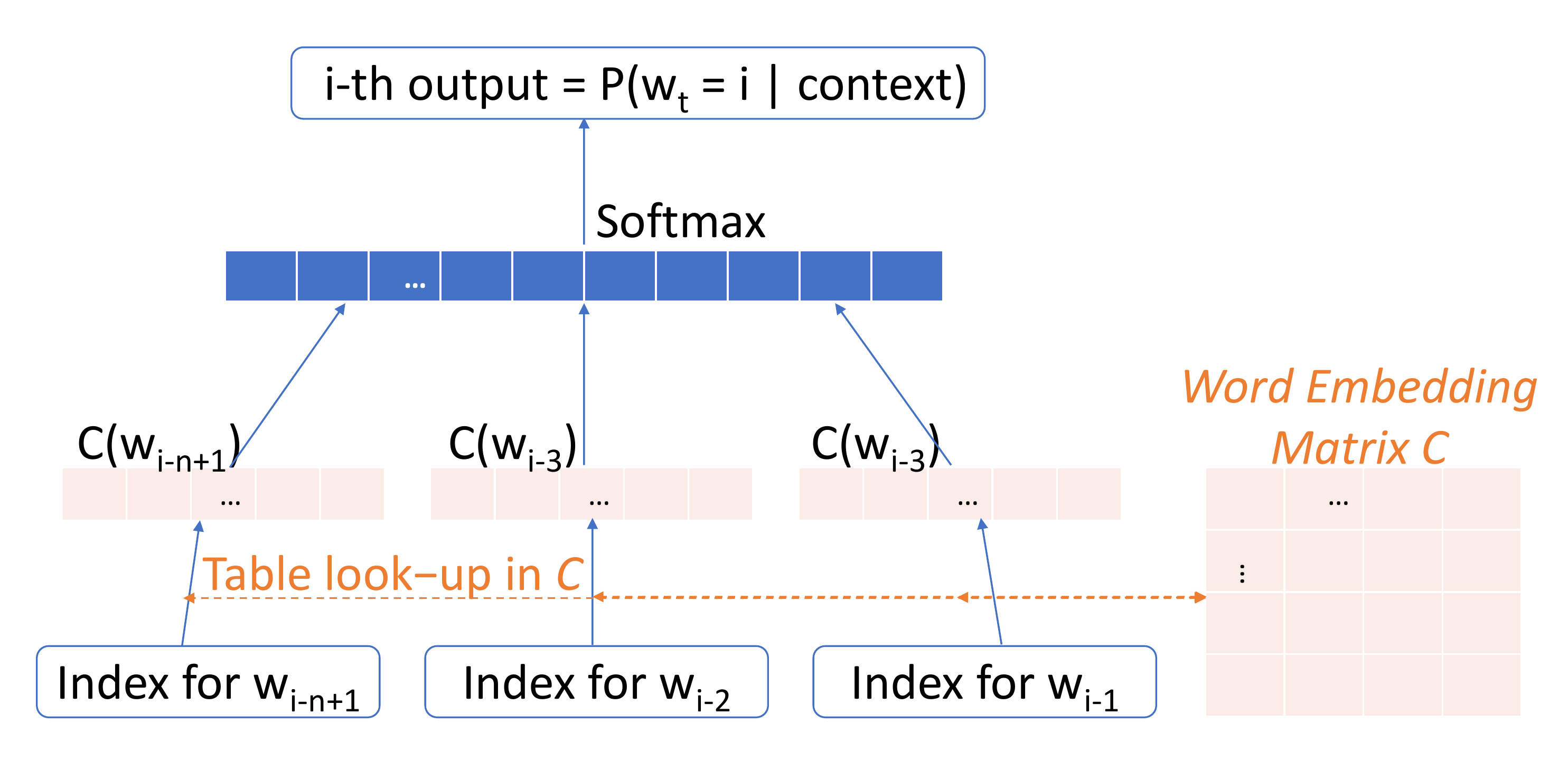}
\caption[Neural architecture]{Neural architecture: $f(i,w_{t-1},\cdots ,w_{t-n+1})=g(i,C(w_{t-1}), \cdots ,C(w_{t-n+1}))$, where $g$ is the neural network and $C(i)$ is the i-th word vector.}
\label{fig:nnet}
\end{figure}
Function $f$ consists of two parts: 1) A mapping $C$ from any element $i$ of $V$ to its associated word vector $C(i) \in R_m$. In practice, $C$ is represented by a $ |V| \times m$ matrix of free parameters. 2) A probability function $g$ that maps an input sequence of word vectors in context $(C(w_{t-n+1} ),\ldots,C(w_{t-1} ))$ to a conditional probability distribution over words in $V$ for the next word $w_t$. The output of $g$ is a vector whose i-th element estimates the probability $\widehat{p}(w_{t=i}|w_{t-n+1}^{t-1} )$ as in Fig~\ref{fig:nnet}. A feedforward or recurrent neural network with parameters $\omega$ may be applied for the function $g$. The parameters of the mapping $C$ are feature vectors themselves, represented by a $|V| \times m$ matrix $C$ whose row $i$ is the word vector $C(i)$ for word $i$. The total set of parameters is $\theta = (C,\omega)$. Training is achieved by looking for $\theta$ which maximizes the log-likelihood of the input dataset:
\begin{equation}
L= \frac{1}{T} \sum_t(log(g(i,C(w_{t-1}),\ldots ,C(w_{t-n+1})))).\end{equation}
\subsection{Network Traffic Generation}
We propose a generative method corresponding to the trained models. Essentially, a new sequence of network activities (represented by cluster IDs) is generated by "gluing" very short and overlapping pieces of network activities that have been seen frequently in the training data. The rules for obtaining the probability of the next network activity depends on the method used (Markov model or neural language model) and are somewhat implicit.

In order to generate network traffic, sequences of activities (clusters) are generated; then flows corresponding to each cluster are sampled from the original traffic, and eventually packets of these flows are transmitted in the order indicated by the generated sequences. In the global sequence aggregation, the sequence model is trained on flows of all of the network members, and correspondingly, a single global sequence is generated that simulates network activities of all network peers, ordered by their time of occurrence. The sequence ends when the estimated end time is reached; therefore, the sequence size is mainly affected by the estimated transmission times which are based on transition time statistics that were extracted from the original traffic. In the IP-based aggregation, a sequence is generated for each pair of IPs (addresses are regenerated), where sequences end when reaching the end state. When using the neural language model, a context of $m$ previous clusters must be updated in each iteration for predicting the next cluster; when using the Markov model the previous cluster is updated in the same manner.

The traffic generation algorithm is outlined in Algorithm ~\ref{algo_ntg}. Its input contains network traffic data ($D$) to be used for reconstructing recorded packets when generating new traffic, as well as both the clustering model ($M_{Clusters}$) and sequence model ($M_{Sequences}$).
\IncMargin{1em}
\begin{algorithm}
\SetKwData{trans}{trans}\SetKwData{bins}{bins}\SetKwData{context}{context}
\SetKwFunction{ExtractFlows}{ExtractFlows}\SetKwFunction{AttachClusters}{AttachClusters}\SetKwFunction{min}{min}\SetKwFunction{max}{max}
\SetKwFunction{FirstTime}{FirstTime}\SetKwFunction{LastTime}{LastTime}
\SetKwFunction{TransitionTimes}{TransitionTimes}\SetKwFunction{ExtractHist}{ExtractHist}
\SetKwFunction{SampleFirstCluster}{SampleFirstCluster}\SetKwFunction{Update}{Update}\SetKwFunction{SampleFlow}{SampleFlow}\SetKwFunction{Sample}{Sample}\SetKwFunction{PredictCluster}{PredictCluster}\SetKwFunction{GetPackets}{GetPackets}\SetKwFunction{GenerateIPs}{GenerateIPs}\SetKwFunction{SetIPs}{SetIPs}
\SetKwInOut{Input}{input}\SetKwInOut{Output}{output}
\Input{A sequence model $M_{Sequences}$,\ a clustering model $M_{Clusters}$,\ and traffic data $D$}
\Output{Generated traffic}
\BlankLine
$D^*\leftarrow$ \ExtractFlows{$D$}\;
$D^*\leftarrow$ \AttachClusters{$D^*, M_{Clusters}$}\;
$t_{start},t_{end} \leftarrow \FirstTime(D),\LastTime(D)$\;
$\trans \leftarrow \ExtractHist(\TransitionTimes(D^*),\bins= \sqrt{|D^*|})$; \tcp{transition times between flows}
\eIf{using global aggregation}
{$c\leftarrow$ \SampleFirstCluster{$M_{Sequences}$}\;
$t\leftarrow t_{start}$\;
 \While{$t < t_{end}$}{
 $f\leftarrow$ \SampleFlow{$D^*$,$c$}\;
 $t\leftarrow t+$ \Sample{\trans}\;
 \context $\leftarrow$ \Update(\context,$c$)\;
 $c\leftarrow$ \PredictCluster{$M_{Sequences}$,\context}\;
 Add \GetPackets{$f$,$D$} to results set\;
}
}
{\tcp{If using IP-based aggregation}
\For{each pair $IP_{src}$,$IP_{dest}$ in $D$}
{
 \context $\leftarrow$ \Update(\context,$start$)\;
 $c\leftarrow$ \PredictCluster{$M_{Sequences}$,\context}\;
 $IP_{src}^*$,$IP_{dest}^*\leftarrow$ \GenerateIPs{}\;
 $t\leftarrow t_{start}$\;
  \While{$c != end$}{
 $f\leftarrow$ \SampleFlow{$D^*$,$c$}\;
 $t\leftarrow t+$ \Sample{\trans}\;
 Add \SetIPs{\GetPackets{$f$,$D$},$IP_{src}^*$,$IP_{dest}^*$} to results set\;
 \context $\leftarrow$ \Update(\context,$c$)\;
 $c\leftarrow$ \PredictCluster{$M_{Sequences}$,\context}\;
}
}
}
\caption{Network traffic generation}\label{algo_ntg}
\end{algorithm}\DecMargin{1em}
\section{Evaluation}
\subsection{Evaluation Methods}
The following will be used to evaluate our proposed framework and compare the methods examined. We use the Silhouette score to evaluate how well did we cluster the flows. 

Many features were extracted from the network traffic and to demonstrate how well the method preserved each of them, we use hypothesis tests to examine whether each feature’s source and synthesized data come from the same distribution. We also used n-gram perplexities to examine whether the distribution of a sequence of feature values was preserved. 

\subsubsection{Silhouette Score}
The silhouette score \cite{Rousseeuw1987} measures how much a member is similar to its cluster (cohesion) compared to other clusters (separation). The score ranges from $-1$ to $+1$, with higher scores indicating that the member is well-grouped (similar to the cluster itself and far from neighboring clusters). High scores for most cluster members indicate that the clustering configuration is appropriate.

The silhouette score for a single flow $f_i$ is calculated as follows:
\begin{equation}
s(i)= \frac{b(f_i )-a(f_i)}{max(a(f_i ),b(f_i))},
\end{equation}
where $a(f_i)$ is the average distance from $f_i$ to all other flows within its cluster, and $b(f_i)$ is the lowest average distance of $f_i$ to all flows in any other cluster (the distance to its second nearest cluster). We used the mean silhouette score of the clustered flows to evaluate the structure of the clustering configuration and tune the number of clusters.
\subsubsection{Two-Sample Kolmogorov-Smirnov Test}
The KS (Kolmogorov-Smirnov) hypothesis test \cite{Massey1951} is a nonparametric test of the equality of continuous, one-dimensional probability distributions, which can be used to compare two samples. It quantifies a distance between the empirical distribution functions of two samples. The null distribution of this statistic is calculated under the null hypothesis that the samples are drawn from the same distribution. We use this test to examine whether a continuous feature has the same distribution in both the original and generated traffic.
\subsubsection{K-Sample Anderson-Darling Test}
AD (Anderson-Darling) hypothesis tests \cite{Stephens1974} are used to determine whether several collections of observations can be estimated as coming from the same population, where the distribution function does not have to be specified. It is a modification of the KS test. We use this test to examine whether a nominal feature has the same distribution in both the original and generated traffic.
\subsubsection{Perplexities}
Perplexity \cite{Brown1992} measures the success of the probability distribution or the probability model to predict a sample. It may be used to compare probability models. Low perplexity indicates that the probability distribution is good at predicting the sample. In natural language processing, perplexity is often used to evaluate language models by comparing n-gram perplexities in the original and generated data. We adapt this idea to examine the preservation of sequences of network activities for each source IP.

The perplexity of a discrete random variable $X$ is defined as:
\begin{equation}
2^{H(X)}=2^{- \sum_xp(x)log_2(p(x))},
\end{equation}
where $H(p)$ is the entropy (in bits) of the distribution over each possible value $x$.
\subsection{The Data}
Experiments were conducted on two network traffic collections.
The first is a 24-hour sample of network traffic at Ben-Gurion University of the Negev's student labs (BGU), and the second is a sample of four hours of traffic from testing laboratories at Dell EMC Corporation (EMC). Various characteristics of these two collections are listed in Table~\ref{tab:data}. As can be seen in the table, the two traffic collections have different characteristics; they have different collection times, and there are almost three times more flows in the EMC traffic, but the flows in the EMC traffic are much shorter than BGU's flows and contain far less packets. In addition, although EMC's traffic contains about half the number of peers compared to BGU's traffic, it contains six times more communicating IP pairs, making the traffic more complicated, at least for the IP-based aggregation. 

\begin{table}[htbp!] 
\centering   
  \caption{Various characteristics of the traffic used for the experiments.}
  \label{tab:data}
  \begin{tabular}{lcc}
    \toprule
      &BGU &EMC\\
    \midrule
    \textbf{Size} &20.1 Gb &2.79 Gb\\
    \textbf{Packets} &19,350,505 &8,700,370\\
    \textbf{Flows} &190,462 &549,350\\
    \textbf{Source IPs} &1,329 &633\\
    \textbf{Destination IPs} &1,309 &435\\
    \textbf{Source \& Dest} &3,322 &24,760\\
    \textbf{Minimal Time} &1.10.2013 1:05:57 &13.6.2017 07:38:27\\
    \textbf{Maximal Time} &2.10.2013 1:11:07 &13.6.2017 12:59:14\\
    \textbf{Sequence Size} &$57 \pm 721$ flows&$22 \pm 367$ flows\\
\bottomrule
\end{tabular}
\end{table}

\subsection{Results and Discussion}
In our first set of experiments, we focused on the evaluation and adaptation of the clustering model. We examined three different clustering configurations by changing the number of clusters for each configuration ($K = 100, 500, 1000$). The BGU preprocessed traffic was clustered according to each of these configurations using scikit-learn's k-means implementation. For each configuration evaluated a sequence model was then trained using the global Markov model and used for generating five samples of perturbed traffic.

\begin{figure}[htbp!] 
\centering   
\includegraphics[width= 0.7\linewidth]{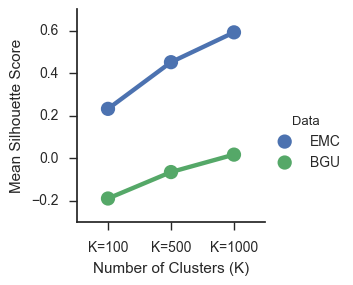}
\caption{Mean silhouette scores versus number of clusters.}
\label{fig:Silhouette}
\end{figure}

\begin{figure*} 
\centering   
\includegraphics[width=0.7\linewidth]{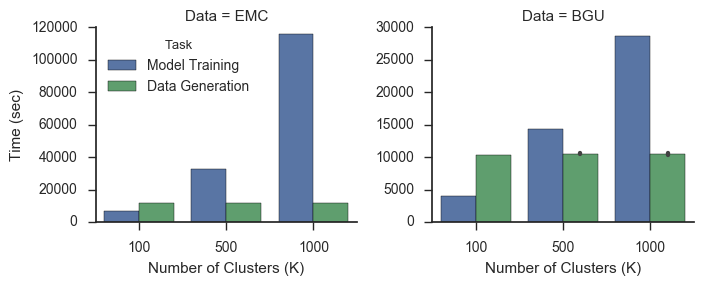}
\caption[The running times versus the number of clusters.]{The running times versus the number of clusters for the model training (blue) and data generation (green) tasks.}
\label{fig:clusterTimes}
\end{figure*}

We assessed the quality of the clusters' structure with the silhouette score, as described in Fig~\ref{fig:Silhouette}. Recall that higher scores indicate a clearer division into clusters, so the results demonstrate that for the configurations examined, increasing the number of clusters ($K$) improves the structure of the clusters, however this is at the price of significantly increasing model training times, as seen in Fig~\ref{fig:clusterTimes}. 

Next, we aimed at understanding which of the features had the greatest effect on the clustering process. For this task we used the Random Forest classifier, which is commonly used for similar tasks \cite{Zhang2015}. For each clustering configuration examined, we trained a Random Forest model to classify the preprocessed traffic into clusters, where the target variable (\textit{clusterID}) is the ID of the cluster as obtained in the clustering phase. We then extracted the top-10 most influencing features in the trained Random Forest model (measured based on the information gain of the various features). Table~\ref{tab:top10} provides information about the top-10 most influencing features for each configuration examined based on the BGU traffic. The top-six features in all configurations are based on the time of day the flow occurred (\textit{day time} feature). The \textit{TCP SeqID} feature is also dominant for 500 and 1000 clusters, while for 100 clusters the \textit{TCP window size} feature is preferred. The day of week in which the flow occurred (\textit{week day} feature) is also influential when using 100 or 500 clusters. Quite similar results were obtained for the EMC dataset.
\begin{table}[htbp!] 
\centering 
\caption[Top-10 most influencing features for clustering the
BGU traffic.]{Top-10 most influencing features for clustering the
BGU traffic, using various numbers of clusters ($K$). (Features extracted from the same attribute appear in the same color.)}
  \label{tab:top10}
  \begin{tabular}{clll}
    \toprule
    Rank &K=100 &K=500 &K=1000\\
    \midrule
    1 &{\color{red}day time last} &{\color{red}day time min} &{\color{red}day time min}\\
    2 &{\color{red}day time min} &{\color{red}day time sum} &{\color{red}day time mean}\\
    3 &{\color{red}day time sum} &{\color{red}day time last} &{\color{red}day time first}\\
    4 &{\color{red}day time first} &{\color{red}day time first} &{\color{red}day time sum}\\
    5 &{\color{red}day time max} &{\color{red}day time max} &{\color{red}day time max}\\
    6 &{\color{red}day time mean} &{\color{red}day time mean} &{\color{red}day time last}\\
    7 &{\color{blue}TCP win size sum} &IP ttl sum &{\color{green}TCP seqID first}\\
    8 &{\color{cyan}week day first} &{\color{green}TCP seqID mean} &{\color{green}TCP seqID sum}\\
    9 &{\color{blue}TCP win size thirdQ} &{\color{cyan}week day sum} &{\color{green}TCP seqID last}\\
    10 &packet size mean &{\color{green}TCP seqID last} &{\color{green}TCP seqID mean}\\
\bottomrule
\end{tabular}
\end{table}

\begin{figure*}[htbp!] 
\centering 
\includegraphics[width=0.8\linewidth]{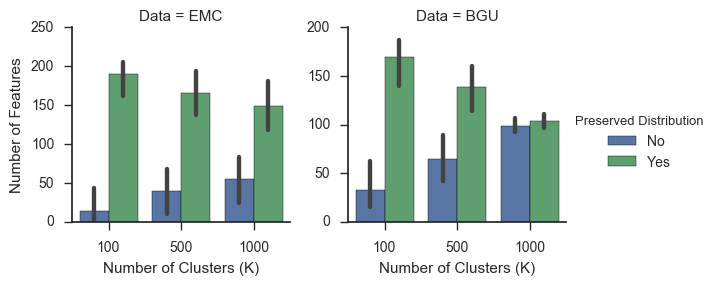}
\caption[The number of features whose distribution is preserved versus the number of clusters.]{The number of features whose distribution is identical (green) or different (blue) in both the original and generated traffic, according to the KS and AD tests ($\alpha=0.01$) versus the number of clusters.}
\label{fig:clusterDist}
\end{figure*}

Another way to evaluate the quality of the clustering results is to check to what extent the probabilities of the different features have been maintained. For each of the clustering configurations examined and for each of the 197 numeric features appearing in the data, we used the two-sample KS test to examine if the distribution of the feature in the original and generated traffic was the same at a significance level of 0.01. In a similar manner, we used the two-sample AD test to examine whether the distribution of seven nominal features was preserved. Fig~\ref{fig:clusterDist} shows the results obtained for each cluster configuration tested. A hundred clusters performed best, preserving the distribution of most of the features.
Using more than 100 clusters weakens the results, probably because rare clusters may not appear in the generated traffic, so that the lack of possible values for members of these clusters leads to a change in statistic, but even with 1000 clusters, at least $50\%$ of the features are preserved.

\begin{figure*}[htbp!] 
\centering   
\includegraphics[width=0.8\linewidth]{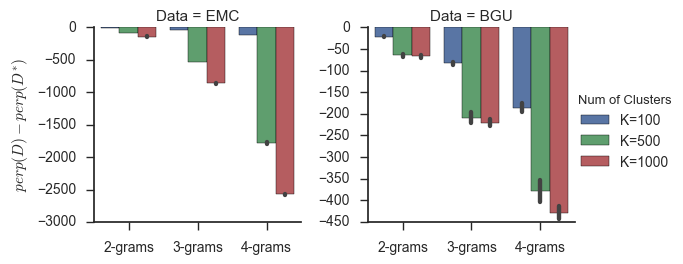}
\caption[The difference between perplexities in the original and generated data for n-grams.]{The difference between perplexities in the original and generated data for n-grams of three possible sizes versus the number of clusters.}
\label{fig:clusterNGrams}
\end{figure*}

In order to estimate how well sequence patterns are preserved in the generated traffic, we measured perplexities for n-grams with three different sizes ($n = 2, 3, 4$) and calculated the differences between perplexities in the original and generated traffic. In general, it is more difficult to preserve sequences for long subsequences, so the perplexities and their differences usually increase with $n$. As shown in Fig~\ref{fig:clusterNGrams}, 100 clusters lead to the lowest difference between n-gram perplexities in the original and generated data, and therefore this number of clusters represents the best clustering configuration for preserving sequence patterns. This is reasonable as reducing the number of parameters makes it easier to preserve the distribution.
Using 100 clusters lead to the best performance in terms of both the quality of the generated traffic and the running time; we therefore chose to set the clustering configuration to 100 clusters.

In the second set of experiments we compared four methods for learning the sequence model and generating network traffic: a Markov model on a global sequence (Global-MM), a Markov model on sequences of IP pairs (IP-MM), a neural language model on a global sequence (Global-NNet), and a
neural language model on sequences of IP pairs (IP-NNet). A fifth method called Random was added for comparison, where flows are randomly sampled from the clustered traffic.

The Markov model was set to accept a single state as input (states are clusterIDs in our case). The neural language model was set to receive a three-state input context (the 3 previous clusterIDs in the sequence). It has an 64-dimensional embedding layer and a softmax activation function, and it was trained for 20 epochs on minibatches (size 256) using the RMSProp optimizer (with a learning rate of 0.001). We preprocessed the two traffic collections and clustered each collection into 100 clusters. Then, for each collection of clustered traffic, we trained a sequence model and generated five network traffic samples using each of the examined methods. In the current study we used a small neural network as a language model to allow faster adjustment of the various parameters. In the future we plan to explore the use of deeper architectures such as recurrent neural networks.

\begin{figure*}[htbp!] 
\centering   
\includegraphics[width=0.8\linewidth]{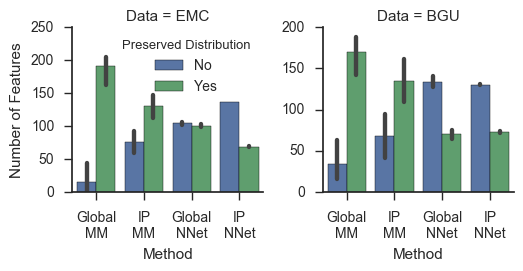}
\caption[The number of features whose distribution preserved using each method.]{The number of features whose distribution is identical (green) or different (blue) in both the original and generated data using each method examined, according to KS and AD tests ($\alpha=0.01$).}
\label{fig:seqDist}
\end{figure*}

We examined how well network traffic generated by the various methods preserved the distribution of the various features. Again, we used the KS and AD tests ($\alpha=0.01$) to determine if the distribution of the numeric / nominal feature (correspondingly) in the original and generated traffic came from the same distribution. The number of features  distribution preserved/unpreserved using each of the examined methods is presented in Fig~\ref{fig:seqDist}. Markov model-based methods better preserve feature distribution, probably because more clusters are included in the generated data. Global methods also preserve feature distribution better, as the IP-based sequence aggregation tends to disrupt the clusters' distribution. The clusters’ distribution is mainly disrupted for the noise inserted by predicting sequence sizes (derived by predicting transfers to the end state). The Global MM method outperformed in this task, preserving distributions of about 100\% of the extracted features.

\begin{figure*}[htbp!] 
\centering  
\includegraphics[width=\linewidth]{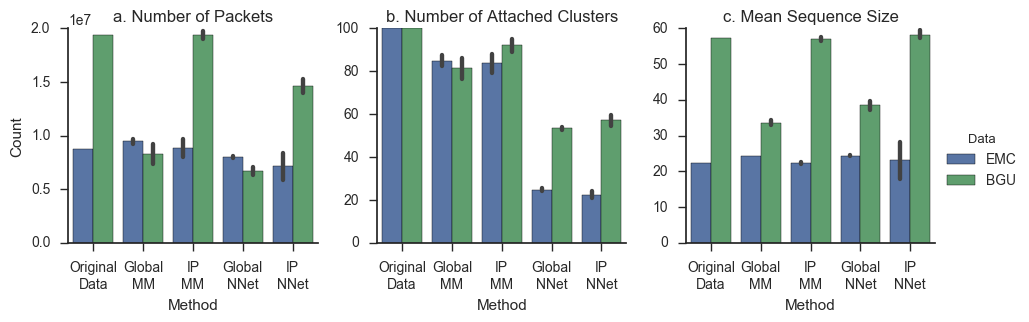}
\caption{(a) Number of packets, (b) number of clusters, and (c) mean sequence size in the original traffic and generated traffic using each of the methods examined.}
\label{fig:seStats}
\end{figure*}

Fig~\ref{fig:seStats}(a) compares the number of packets generated using the four methods and the number of packets in the original traffic. In the EMC traffic all of the methods produced similar data volumes, but in the BGU data the IP-based methods performed better than the global methods. This seems to be caused by the stopping condition of reaching the end time, which was defined for the global generation algorithm. As transmission times in the BGU data do not follow a simple pattern of fixed intervals, disrupted times were generated according to the extracted statistics, accelerating the progress to the end time. This early stopping also shortened the length of sequences generated for the BGU data with the global methods (Fig~\ref{fig:seStats}(c)). The number of clusters in the original and generated traffic is presented in Fig~\ref{fig:seStats}(b). The neural language model-based methods seldom reach rare clusters; therefore, they produce traffic from fewer clusters than those generated using Markov model methods.

\begin{figure*}[htbp!]  
\centering  
\includegraphics[width=\linewidth]{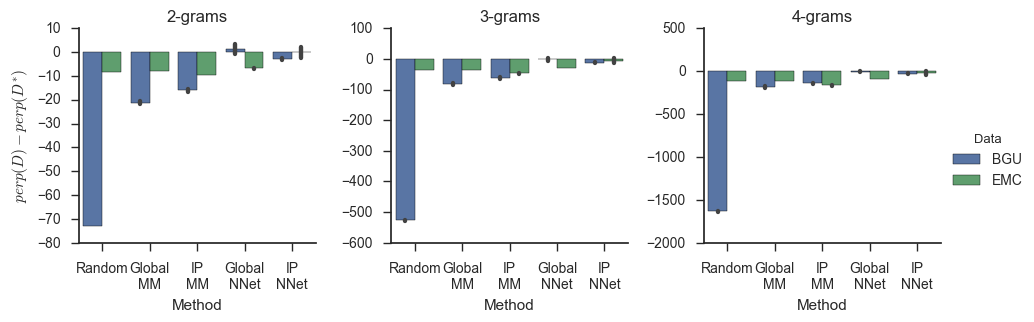}
\caption[Preserving sequence patterns with various methods.]{Preserving sequence patterns with various methods: the difference between perplexities in the original and generated traffic for n-grams of three possible sizes.}
\label{fig:seqNGram}
\end{figure*}

We evaluated the preservation of sequence patterns by calculating perplexities for n-grams of different sizes ($n = 2, 3, 4$) in the original and generated data. We calculated the difference between the perplexities; the smaller the difference, the better the n-gram's distribution was preserved. As can be seen in Fig~\ref{fig:seqNGram}, in the BGU traffic, all of the methods examined outperform the Random approach; neural language model-based methods significantly improved the preservation of n-grams in all examined sizes, with a slight advantage for IP-based compared to global sequence aggregation. In the EMC traffic, however, the Markov model-based methods have similar results to that of the Random approach. It appears that in the EMC traffic most of the sequences are concentrated in a few dominant clusters, so that simply maintaining clusters' distribution like the Random approach does is enough to provide similar n-gram perplexities. Using neural language model-based methods, however, preserves n-gram perplexities even better, with almost identical n-gram perplexities in the original and generated traffic.
\section{Conclusions}
In this study, we presented the NTG framework for generating network traffic, which preserves both the distribution of multiple traffic characteristics and sequence patterns at the flow level. This method is particularly powerful for diverse network traffic (as observed in the BGU data); in this scenario, patterns that an analyst will find difficult to extract and encode manually due to their complex nature are successfully extracted by the NTG framework. Experiments conducted on two real network traffic collections demonstrated that the traffic generated with the NTG framework preserves multiple characteristics of the original traffic as well as sequence patterns of network activities.

This study demonstrates that it is possible to use a limited amount of available network traffic to create scalable, adjustable, and representative network traffic. The traffic generated can be used for research or system development. The results obtained show that there is a trade-off between preserving distributions of basic traffic characteristics and preserving sequences of network activities. The use of a neural language model that strengthens sequence pattern preservation led to a reduction in the number of features whose distribution was preserved. Preserving sequences of IP pairs (IP-based aggregation), instead of preserving the traffic as a single sequence (global aggregation), improved the preservation of network activity sequences when using the Markov model, but this was again at the expense of the preservation of basic traffic characteristics. In the future, we plan to examine the possibility of strengthening the methods that performed best in preserving sequence patterns in our current experiments, so that the negative consequences to maintaining the distribution of basic traffic characteristics will be mitigated. We believe that extracting sequence patterns of network activities can contribute significantly to the identification of network attacks, and we plan to examine this direction in the future and develop an anomaly detection method based on the NTG framework’s extracted model.

\section{Acknowledgments}
We would like to thank the Dell EMC Cyber Solutions Group for sharing their knowledge and capturing real network traffic for our experiments. Special thanks go to Boris Giterman who guided the project on behalf of Dell EMC and enabled this research to reach its destination by constantly striving for fruitful cooperation. This research was funded by the Israel Innovation Authority under the MAGNETON Program.

\bibliographystyle{unsrt}  
\bibliography{template}  


\end{document}